\documentclass[aps,nofootinbib,showpacs,preprintnumbers,amsmath,amssymb]{revtex4}
\usepackage{epsfig}
\usepackage{epsf}
\usepackage{amssymb}
\usepackage{amsmath}

\begin{document}
\preprint{USM-TH-179}

\title{An approach for evaluation of observables in analytic versions of QCD}\footnote{
hep-ph/0601050v3 - one reference added; no other changes;
version to appear in J. Phys. G.}

\author{Gorazd Cveti\v{c}}
\author{Cristi\'an Valenzuela}
\affiliation{Dept.~of Physics, Universidad T\'ecnica
Federico Santa Mar\'{\i}a, Valpara\'{\i}so, Chile} 

\date{\today}

\begin{abstract}
We present two variants of an approach for evaluation of observables 
in analytic QCD models. The approach is motivated by the skeleton 
expansion in a certain class of schemes. We then evaluate the Adler function 
at low energies in one variant of this approach, in various analytic QCD models 
for the coupling parameter, and compare with perturbative QCD 
predictions and the experimental results. We introduce two analytic QCD models
for the coupling parameter which reproduce the measured value of the
semihadronic $\tau$ decay ratio. Further, we evaluate the Bjorken polarized sum 
rule at low energies in both variants of the evaluation approach, using for the 
coupling parameter the analytic QCD model of Shirkov and Solovtsov, 
and compare with values obtained by the evaluation approach 
of Milton {\em et al.\/} and Shirkov.

\end{abstract}

\maketitle

Consider an observable $\mathcal{O}(Q^2)$ depending on a single 
space-like scale $Q^2 (\equiv - q^2) >0$ and assume that the 
skeleton expansion for this observable exists:
\begin{eqnarray}
\mathcal{O}_{\text{skel}}(Q^2) &=&
\int_0^\infty \frac{dt}{t}\: F_{\mathcal{O}}^{\mathcal{A}}(t) \: 
a_{\text{pt}}(t e^C Q^2)   
+ \sum_{n=2}^{\infty} s_{n-1}^\mathcal{O} 
\left[ \prod_{j=1}^{n} \!\int_0^{\infty}\!\frac{d t_j}{t_j} 
a_{\text{pt}}(t_j e^C Q^2) \right]
F_{\mathcal{O}}^{\mathcal{A}}(t_1,\!\ldots\!,t_n).
\label{sk1}
\end{eqnarray}
The observable is normalized such that
$\mathcal{O}(Q^2)= a_{\text{pt}}$ at first order in perturbation theory.
The characteristic functions $F_{\mathcal{O}}^{\mathcal{A}}$ 
are symmetric functions and have the following normalization:
\begin{equation}
\int_0^\infty \frac{dt}{t} \ F_{\mathcal{O}}^{\mathcal{A}}(t)=1,
\qquad
\int \frac{dt_1}{t_1}\frac{dt_2}{t_2} \ 
F_{\mathcal{O}}^{\mathcal{A}}(t_1,t_2)=1, \  \dots,
\label{mom0}
\end{equation}
and $s_i^\mathcal{O}$ are the skeleton coefficients.
The perturbative running coupling 
$a_{\text{pt}}(Q^2)\equiv \alpha(Q^2)/\pi$ obeys the 
renormalization group (RG) equation:
\begin{equation}
\frac{\partial a_{\text{pt}}(Q^2)}{\partial\log Q^2}
= -[\,\beta_0 a_{\text{pt}}^2(Q^2) + 
\beta_1 a_{\text{pt}}^3(Q^2)+\dots] \ .
\label{rg}
\end{equation}
In QCD, the first two coefficients
$\beta_0= (1/4)(11-2 n_f/3)$ and 
$\beta_1=(1/16)(102-38 n_f/3)$ 
are scheme-independent in mass-independent schemes;
$n_f$ is the number of active quarks flavors.
The value of $C$ depends on the value of the scale $\Lambda$ 
in $a_{\text{pt}}$ 
($\Lambda^2_{(C)} = \Lambda^2_{(0)} e^C$) \cite{Neubert}. 
In ${\overline {\rm MS}}$ scheme $C={\overline C} \equiv -5/3$. 
The skeleton integrands and integrals are independent of $C$.
Expansion (\ref{sk1}) exists in QED if one excludes 
light-by-light subdiagrams \cite{Bjorken:1979dk,Brodsky}.
In the QCD case, the leading skeleton part
was investigated in Refs.~\cite{Neubert,Gardi:1999dq}.
We will assume that expansion (\ref{sk1}) exists in a 
certain class of schemes.

On the other hand, the RG-improved perturbation expansion for the 
observable $\mathcal{O}(Q^2)$ is given by
\begin{equation}
\mathcal{O}_{\text{pt}}(Q^2)= a_{\text{pt}}(Q^2)
+ \sum_{n=2}^{\infty} c_{n-1} a_{\text{pt}}^{n}(Q^2) \ .
\label{pts}
\end{equation}
Expanding $a_{\text{pt}}(t e^C Q^2)$ around $t=e^{-C}$ inside
the integrals in Eq.~(\ref{sk1}) must give Eq.~(\ref{pts}).

The skeleton expansion is a reorganization of the perturbation series 
such that each term in (\ref{sk1}) corresponds to the sum of an 
infinite number of Feynman diagrams. These sums, however, do not 
converge. Although it is possible to assign a value to these sums, 
this value is not unique, a renormalon ambiguity is present. 
In formulation (\ref{sk1}) the ambiguities arise 
from the (nonphysical) Landau singularities of 
$a_{\text{pt}}(t e^C Q^2)$ in the non-perturbative 
space-like region $0 < Q^2 \leq \Lambda^2$ \cite{Neubert}.
The difference between possible integration paths (prescriptions) 
is a measure of the size of the renormalon ambiguity.

The perturbative coupling is a solution of the 
n-loop RG equation (\ref{rg}).
It can be found iteratively for $Q^2 \gg \Lambda^2$
\begin{equation}
a_{\text{pt}}(Q^2)=
\sum_{i=1}^n \sum_{j=0}^{i-1} k_{ij}\:
\frac{(\log L)^j}{L^i},
\label{apt}
\end{equation}
where $L=\log(Q^2/\Lambda^2)$
and $k_{ij}$ are constants depending on the $\beta$-function 
coefficients.
At energies $Q \alt 1$ GeV the perturbative 
result (\ref{pts}),  (\ref{apt}) is not reliable.
At these energies $a_{\text{pt}}$ starts being dominated by the 
Landau singularities at $0 < Q^2 \leq \Lambda^2$.
These singularities are a consequence of the perturbative RG
Eq.~(\ref{rg}) and are located in the region where this
equation is not valid.
Furthermore, from general arguments (causality) \cite{ShS} 
one concludes that the coupling parameter must be analytic in 
the whole $Q^2$-plane excluding the time-like (Minkowskian) semiaxis.
On this semiaxis, singularities associated with 
asymptotic states appear.

With this motivation it seems reasonable to replace the perturbative 
coupling by a new coupling $\mathcal{A}_1(Q^2)$ differing from 
$a_{\text{pt}}(Q^2)$ significantly only in the non-perturbative 
region and having the required analyticity properties.
This replacement is not unique and should be considered as a 
phenomenological model.
Using a dispersion relation, Shirkov and Solovtsov \cite{ShS} 
proposed the following replacement:
\begin{eqnarray}\label{MAa1}
a_{\text{pt}}(Q^2)&\equiv&
\frac{1}{\pi}\int_{-\Lambda^2}^\infty 
\frac{d\sigma}{\sigma+Q^2}\: \rho_1(\sigma)
\: \: 
\mapsto
\: \:
\mathcal{A}_1^{\text{(MA)}}(Q^2) =
\frac{1}{\pi}\int_0^\infty \frac{d\sigma}{\sigma+Q^2}\: \rho_1(\sigma),
\end{eqnarray}
where $\rho_1=\text{Im}[a_{\text{pt}}(-\sigma-i\epsilon)]$,
and $a_{\text{pt}}$ is, e.g., given by Eq.~(\ref{apt}).\footnote{
An efficient method for evaluation of $\mathcal{A}_1^{\text{(MA)}}$
was developed in Ref.~\cite{Alekseev:2002zn}.
A different evaluation of $a_{\rm pt}$ and $\rho_1$
was presented in Ref.~\cite{Kurashev:2003pt}.} 
We will refer to this as the minimal analytic (MA) procedure.
In MA the discontinuity of the analytic coupling along 
the Minkowskian semiaxis is by construction the same as the one 
of $a_{\text{pt}}$.\footnote{
Other analytization procedures of $a_{\rm pt}$
focus on the analyticity properties of the
beta function \cite{Nesterenko,Raczka}.}
Below we shall consider generalizations 
of this analytization procedure.

Once one analytizes $a_{\text{pt}}(Q^2)$ using Eq.~(\ref{MAa1}) or 
other procedure, the question arises how to treat a known
truncated perturbation series (TPS). For perturbation series (\ref{pts}) 
there is no unique way to analytize higher powers of $a_{\text{pt}}$.
One possibility is to apply the MA procedure to each 
power of $a_{\text{pt}}$ \cite{Milton:1997mi}:
\begin{equation}
a_{\text{pt}}^k(Q^2) \: \mapsto \:
\mathcal{A}_k^{\text{(MA)}}(Q^2)=
\frac{1}{\pi}\int_0^\infty \frac{d\sigma}{\sigma+Q^2}\: \rho_k(\sigma)
\qquad (k=1,2,\ldots) \ ,
\label{MAak}
\end{equation}
where $\rho_k=\text{Im}[a_{\text{pt}}^k(-\sigma-i\epsilon)]$
and $a_{\text{pt}}$, e.g., is given by Eq.~(\ref{apt}).\footnote{
An extension of Eq.~(\ref{MAak}) to noninteger powers
was developed in Refs.~\cite{Bakulev:2005gw}.}
Other choices could be, e.g.,  
$a_{\text{pt}}^k \mapsto \mathcal{A}_1^k, \mathcal{A}_1^{k-2} \mathcal{A}_2$,
etc.
In this paper we propose a method to analytize a TPS
based on the skeleton expansion (\ref{sk1}), 
in any chosen version of QCD with
analytic $\mathcal{A}_1(Q^2)$. In Eq.~(\ref{sk1}) the replacement
$a_{\rm pt} \mapsto \mathcal{A}_1$ is made, making the skeleton
expansion terms well-defined integrals.
Next, we Taylor-expand each $\mathcal{A}_1(t_j e^C Q^2)$ there 
around a specific $\ln Q_*^2 = \ln (t_* e^C Q^2)$. In these
expansions, we denote
\begin{equation}
{\widetilde {\mathcal{A}_n}}(Q_*^2) \equiv
\frac{ (-1)^{n-1}}{\beta_0^{n-1} (n-1)!} 
\frac{ \partial^{n-1} {\mathcal{A}_1}(Q_*^2)}{\partial (\ln Q_*^2)^{n-1}} \ ,
\qquad (n=2,3,\ldots)
\label{tAn}
\end{equation}
If we know in expansion (\ref{pts}) TPS with
$n_{\rm max}=3$ (i.e., $c_1$ and $c_2$), it is convenient to
introduce the analytic couplings ${\mathcal{A}_2}$ and
${\mathcal{A}_3}$ according to
\begin{equation}
\begin{split}
{\widetilde {\mathcal{A}_2}}(Q_*^2) & = {\mathcal{A}_2}(Q_*^2) +
\frac{\beta_1}{\beta_0} {\mathcal{A}_3}(Q_*^2) \ , 
\\
{\widetilde {\mathcal{A}_3}}(Q_*^2) & = {\mathcal{A}_3}(Q_*^2) \ .
\end{split}
\label{A2A3}
\end{equation}
When replacing here $\mathcal{A}_k \mapsto a_{\text{pt}}^k$, we obtain
the corresponding truncated RG equations of perturbative QCD (pQCD).
Thus, once one chooses a particular analytic coupling $\mathcal{A}_1$, 
the functions $\mathcal{A}_k$ with $k\geq 2$ 
are defined by Eqs.~(\ref{A2A3}),
or by a higher-$n_{\rm max}$ version thereof. 
In MA model (\ref{MAa1}), the results 
(\ref{MAak}) and (\ref{A2A3}) merge 
when $n_{\rm max}$ increases \cite{CV2} (cf.~also \cite{Kurashev:2003pt}).
In our approach, the basic set of functions is:
$({\mathcal{A}_1}$ and its derivatives 
${\widetilde {\mathcal{A}_2}}, {\widetilde {\mathcal{A}_3}},\ldots)$. 
The set $({\mathcal{A}_1},{\mathcal{A}_2},{\mathcal{A}_3},\ldots)$
was introduced for the convenience of comparison with pQCD.

The afore-mentioned Taylor-expansion around $\ln Q_*^2 = \ln (t_* e^C Q^2)$ 
for the order $n_{\rm max}=3$ gives
\begin{eqnarray}
\mathcal{A}_1(t e^C Q^2) &\approx& 
\mathcal{A}_1(Q_*^2) - \beta_0 \ln (t/t_*) {\widetilde {\mathcal{A}_2}}(Q_*^2) +
\beta_0^2 \ln^2(t/t_*) {\widetilde {\mathcal{A}_3}}(Q_*^2)
\label{A1exp1}
\\
& = &
\mathcal{A}_1(Q_*^2)
- \beta_0 \ln(t/t_*)\, \mathcal{A}_2(Q_*^2) 
+ [\beta_0^2 \ln^2(t/t_*)-\beta_1 \ln(t/t_*)] 
\mathcal{A}_3(Q_*^2).
\label{A1exp2}
\end{eqnarray}
Keeping terms corresponding to the third-order approximation we 
obtain a truncated analytic version of $\mathcal{O}(Q^2)$:
\begin{eqnarray}
\mathcal{O}_{\text{tr}}^{\text{(an)}}(Q^2)&=&
\mathcal{A}_1(Q_*^2)
+ \left[ \beta_0 f_1^\mathcal{O}(t_*) \mathcal{A}_2(Q_*^2)+ 
s_1^\mathcal{O}\mathcal{A}_1^2(Q_*^2) \right] 
\nonumber\\
&&
+ \left[
(\beta_0^2 f_2^\mathcal{O}(t_*)+ \beta_1 f_1^\mathcal{O}(t_*))
\mathcal{A}_3(Q_*^2)
+  2 s_1^\mathcal{O} \beta_0 f_{1,0}^\mathcal{O}(t_*) 
\mathcal{A}_1(Q_*^2)\mathcal{A}_2(Q_*^2)
+ s_2^\mathcal{O} \mathcal{A}_1^3(Q_*^2) \right] \ ,
\label{Oantr}
\end{eqnarray}
where $Q_*^2 \equiv t_* e^C Q^2$ and the momenta are
\begin{equation}
\begin{split}
f_i^\mathcal{O}(t_*) 
&=
\int_0^\infty \frac{dt}{t}
  F_{\mathcal{O}}^{\mathcal{A}}(t)(-\log t/t_*)^{i},
\\
f_{i,j}^\mathcal{O}(t_*) 
&= 
\int \frac{dt_1}{t_1}\frac{dt_2}{t_2}
  F_{\mathcal{O}}^{\mathcal{A}}(t_1,t_2)
  (-\log t_1/t_*)^{i}(-\log t_2/t_*)^{j}.
\end{split}
\label{mom1}
\end{equation}
Equation (\ref{Oantr}) is the result of the proposed analytization 
procedure for the series (\ref{pts}) truncated at $\sim a_{\rm pt}^3$.
In the perturbative region one has 
$\mathcal{A}_k\approx a_{\text{pt}}^k$ and Eqs.~(\ref{Oantr}) and 
(\ref{pts}) merge.
If $\mathcal{A}_1(Q^2)$ is well behaved at the origin then all 
$\mathcal{A}_k(Q^2)$ ($k \geq 2$) vanish at this point.
This follows from the (truncated) RG-like Eqs.~(\ref{A2A3}).
Comparison between Eqs.~(\ref{Oantr}) and (\ref{pts}) gives
\begin{equation}
\begin{split}
c_1 &= \beta_0 f_1^\mathcal{O}(e^{-C}) + s_1^\mathcal{O},
\\
c_2 &= \beta_0^2 f_2^\mathcal{O}(e^{-C})\!+\! 
\beta_1 f_1^\mathcal{O}(e^{-C}) \!+\!
  2 \beta_0 s_1^\mathcal{O} f_{1,0}^\mathcal{O}(e^{-C}) + 
s_2^\mathcal{O}.
\end{split}
\label{cj}
\end{equation}
We assume that the skeleton expansion coefficients
$s_j^\mathcal{O}$ and characteristic functions
$F_{\mathcal{O}}^{\mathcal{A}}(t_1, \ldots, t_j)$
are $n_f$-independent when $C$ is $n_f$-independent.
Consequently, in the class of schemes where the coefficients $c_j$ 
of expansion (\ref{pts}) are
polynomials in $n_f$ ($\leftrightarrow \beta_0$)
of order $j$,
relations (\ref{cj}) give us
coefficients $s_1^\mathcal{O}$ and $s_2^\mathcal{O}$
and momenta $f_1^\mathcal{O}$, $f_2^\mathcal{O}$
and $f_{1,0}^\mathcal{O}$. We shall consider observables
for which $c_1$ and $c_2$ are known.
This approach can be continued to higher orders.
The afore-mentioned class of schemes
is parametrized by the RG $\beta_j$ coefficients
($j \geq 2$) which are polynomials in $n_f$ ($\leftrightarrow \beta_0$)
of order $j$ such that
$\beta_j = b_{j0} + b_{j1} \beta_0 + \cdots
b_{jj} \beta_0^j$, where $b_{j0} = {\overline b}_{j0}$
of ${\overline {\rm MS}}$ scheme
and $b_{jk}$ ($k \geq 1$) are the free scheme parameters.

However, the knowledge of the perturbation coefficients 
$c_j$ by itself is not enough 
to obtain the higher-order coefficients which are
not included in Eq.~(\ref{Oantr}). At fourth-order,
the coefficients at $\mathcal{A}_2^2$ and
$\mathcal{A}_1 \mathcal{A}_3$ cannot be obtained
without certain assumptions for the characteristic function
$F_{\mathcal{O}}^{\mathcal{A}}(t_1,t_2)$ \cite{CV2}. 

The leading characteristic function
$F_{\mathcal{O}}^{\mathcal{A}}(t)$ is known 
for many observables on the basis of their all-order large-$n_f$ 
($\leftrightarrow$ large-$\beta_0$) 
perturbation expansion \cite{Neubert,Beneke:1992ch}. 
Therefore, we propose to keep
the leading skeleton term unexpanded, but to expand
the other terms as in Eq.~(\ref{Oantr})
\begin{eqnarray}
\mathcal{O}^{(\rm{an})}_{\text{skel}}(Q^2) =
\int_0^\infty \frac{dt}{t}\: F_{\mathcal{O}}^{\mathcal{A}}(t) \: 
\mathcal{A}_1(t e^C Q^2) + 
s_1^\mathcal{O} \mathcal{A}_1^2(Q_2^2) 
+ \left[ s_2^\mathcal{O} \mathcal{A}_1^3(Q_3^2)
+ 2 s_1^\mathcal{O} \beta_0 f_{1,0}^\mathcal{O}(t^{(2)}) 
\mathcal{A}_1(Q_2^2)\mathcal{A}_2(Q_2^2) 
\right],
\label{sk2}
\end{eqnarray}
where we used two different expansion scales for the NL and NNL skeleton
terms: $Q_2^2 \equiv t^{(2)} e^C Q^2$ and 
$Q_3^2 \equiv t^{(3)} e^C Q^2$, respectively. 
Since $f_{1,0}^\mathcal{O}(t^{(2)}) = f_{1,0}^\mathcal{O}(e^{-C}) +
\ln t^{(2)} + C$ and $f_{1,0}^\mathcal{O}(e^{-C})$ is known, 
it is convenient to use a scale of the
BLM type \cite{Brodsky:1982gc,Brodsky}: 
$t^{(2)} = t^{(2)}_{\ast} \equiv \exp(- C - f_{1,0}^\mathcal{O}(e^{-C}))$
such that $f_{1,0}^\mathcal{O}(t^{(2)}_{\ast})=0$. Consequently,
the $\mathcal{A}_1 \mathcal{A}_2$ term in Eq.~(\ref{sk2})
disappears. Further, the scheme dependence at this level
shows up as the dependence of $s_2^\mathcal{O}$ and $t^{(2)}_{\ast}$ 
on $b_{21}$ and $b_{22}$, respectively \cite{CV2}. 
This allows us to fix the latter
two coefficients, for each specific observable, in such a way
that $s_2^\mathcal{O} = 0$ and, e.g., $t^{(2)}_{\ast} = 1$.
For example, if the starting scheme is ${\overline {\rm MS}}$
(${\overline b}_{21}$, ${\overline b}_{22}$, and
${\overline C} \equiv - 5/3$), the new 
scheme coefficients $b_{2j}$ are
\begin{equation} \label{bij}
\begin{split}
b_{21} &=  {\overline b}_{21} + {\overline c}_{20} + 
\frac{107}{16} {\overline c}_{11} ,\\
b_{22} &=  {\overline b}_{22} + {\overline c}_{21} 
- \frac{19}{4} {\overline c}_{11} + 2 {\overline C} \ {\overline c}_{10} ,
\end{split}
\end{equation} 
where ${\overline c}_{jk}$ are expansion coefficients
of the perturbation coefficient $c_j$,
Eq.~(\ref{pts}), in powers of $\beta_0$, 
in ${\overline {\rm MS}}$ scheme: 
${\overline c}_j = \sum_0^j {\overline c}_{jk} \beta_0^k$.
In the scheme (\ref{bij}),
the skeleton-based expansion (\ref{sk2}) reduces to
\begin{eqnarray}
\mathcal{O}^{(\rm{an})}_{\text{v1}}(Q^2) &=&
\int_0^\infty \frac{dt}{t}\: F_{\mathcal{O}}^{\mathcal{A}}(t) \: 
\mathcal{A}_1(t e^C Q^2) 
+ s_1^\mathcal{O} \mathcal{A}_1^2(e^C Q^2) + 
\mathcal{O}_n \ ,
\label{v1}
\end{eqnarray}
where $\mathcal{O}_n = \mathcal{O}_4$ are now 
formally terms of fourth 
order ($\sim \mathcal{A}_1^4, \mathcal{A}_1^2 \mathcal{A}_2, \ldots$). 
We will call formula (\ref{v1}) the first variant
(``v1'') of our skeleton-motivated evaluation approach.
Adopting the scheme (\ref{bij}), or higher order
generalizations of it, higher order contributions are
absorbed in the two terms of Eq.~(\ref{v1}). 
This scheme-fixing method is particularly useful at
low energies where scheme dependence is important. 

The skeleton QCD expansion, if it exists, is probably valid
only in a specific (yet unknown) ``skeleton'' scheme 
\cite{Gardi:1999dq,Brodsky}.
A possible difference between the latter and the schemes
used here will result in a difference in the evaluation
of the observable ${\mathcal O}(Q^2)$. This
difference, when re-expanded in $a_{\rm pt}$, 
is at most $\sim a_{\rm pt}^4$ subleading-$\beta_0$
(i.e., $\sim \beta_0^2 a_{\rm pt}^4$).

The derivation up until now allows us to present
yet another variant (``v2'') of the evaluation approach,
by keeping the scheme (\ref{bij}) and 
simply replacing $\mathcal{A}_1^2(e^C Q^2)$ by
$\mathcal{A}_2(e^C Q^2)$ in Eq.~(\ref{v1})
\begin{eqnarray}
\mathcal{O}^{(\rm{an})}_{\text{v2}}(Q^2) &=&
\int_0^\infty \frac{dt}{t}\: F_{\mathcal{O}}^{\mathcal{A}}(t) \: 
\mathcal{A}_1(t e^C Q^2) 
+ s_1^\mathcal{O} \mathcal{A}_2(e^C Q^2) + 
\mathcal{O}_n \ ,
\label{v2}
\end{eqnarray}
This formula can be obtained by repeating the
previous derivation, but starting with the skeleton
expansion (\ref{sk1}) without the analytization replacements
$a_{\rm pt} \mapsto \mathcal{A}_1$ there. All the
expansions are then obtained as previously,
but with $a_{\rm pt}^n$ instead of  
$\mathcal{A}_n$, $\mathcal{A}_{n-1}\mathcal{A}_1$, etc.
In this variant, the analytization is performed
at the end, by replacing $a_{\rm pt} \mapsto \mathcal{A}_1$
in the leading-skeleton integral, and replacing
$a_{\rm pt}^2 \mapsto \mathcal{A}_2$ in the term
proportional to $s_1^\mathcal{O}$, leading to Eq.~(\ref{v2}).

We wish to stress that neither variant of the evaluation
approach relies on the existence of the 
skeleton expansion. Our derivation can
be interpreted in the following alternative way:
The formal skeleton expansion (\ref{sk1}) provides us with the tools
to separate the perturbation series of the observable
into several perturbation subseries. The first subseries 
(from the leading skeleton term) includes all the
leading-$\beta_0$ terms, the second subseries (from the
subleading skeleton term) includes all the
leading-$\beta_0$ terms of the rest, etc.
Each of these perturbation
subseries is renormalization scale invariant.
A specific renormalization scheme ($\beta_2, \beta_3, \ldots$)
is then found such that all the perturbation subseries
vanish, except the first two. In the end, the
analytization of the two surviving subseries is performed.

If the perturbation coefficient $c_3$ is known, then the entire 
described procedure can be carried out to one higher order,
i.e., the $\beta_3$-coefficients $b_{3j}$ ($j=1,2,3$) can be
determined so that in Eqs.~(\ref{v1})-(\ref{v2})
$\mathcal{O}_n = \mathcal{O}_5$ 
($\sim \mathcal{A}_1^5, \mathcal{A}_1^3 \mathcal{A}_2, \ldots$),
under certain assumptions for the function 
$F_{\mathcal{O}}^{\mathcal{A}}(t_1,t_2)$ \cite{CV2}.
For example, for the massless Adler function, the 
${\overline c}_3$ coefficient has been estimated as a polynomial in $n_f$
to a high degree of accuracy \cite{Baikov:2002uw},
and the scheme can be found
such that in Eqs.~(\ref{v1})-(\ref{v2}) $\mathcal{O}_n = \mathcal{O}_5$.
Of course, the higher order analytic couplings $\mathcal{A}_n$
($n \geq 2$) are defined in this case by the
$n_{\rm max}=4$ extension of Eqs.~(\ref{A2A3})
\begin{equation}
\begin{split}
{\widetilde {\mathcal{A}_2}}(Q_*^2) & = {\mathcal{A}_2}(Q_*^2) +
\frac{\beta_1}{\beta_0} {\mathcal{A}_3}(Q_*^2) +
\frac{\beta_2}{\beta_0} {\mathcal{A}_4}(Q_*^2) \ , 
\\
{\widetilde {\mathcal{A}_3}}(Q_*^2) & = {\mathcal{A}_3}(Q_*^2) 
+ \frac{5}{2} \frac{\beta_1}{\beta_0} {\mathcal{A}_4}(Q_*^2) \ , 
\\
{\widetilde {\mathcal{A}_4}}(Q_*^2) & = {\mathcal{A}_4}(Q_*^2) \ , 
\end{split}
\label{A2A3A4}
\end{equation}
and expansion of $\mathcal{A}_1(t e^C Q^2)$ is now performed
up to and including ${\widetilde {\mathcal{A}_4}}(Q_*^2)$,
in contrast to Eq.~(\ref{A1exp1}).

In practical evaluations, the form of the analytic coupling
parameter $\mathcal{A}_1(Q^2)$ has to be specified. The most
straightforward is the minimal analytic (MA) coupling (\ref{MAa1}).
The latter model gives the value 
${\overline \Lambda} \equiv {\Lambda}_{(C=-5/3)} \approx 0.4$ GeV
(in ${\overline {\rm MS}}$ and with $n_f=3$)
from fitting high energy QCD observables \cite{Sh}. 
However, in order to reproduce the measured value of the
semihadronic tau decay ratio $r_{\tau}$,
it requires introduction of heavy first generation quark masses 
$m_u \approx m_d \approx 0.25$ GeV \cite{Milton:1997mi}.
Another possibility
would be to modify the MA-coupling at low energies,
e.g., in the following manner
\begin{eqnarray}
\mathcal{A}_1^{(M1)}(Q^2) =
c_f \frac{M_r^2 Q^2}{ (Q^2 + M_r^2)^2} + k_0 \frac{M_0^2}{(Q^2 + M_0^2)}
+ \frac{Q^2}{(Q^2 + M_0^2)} \frac{1}{\pi} \int_{\sigma= M_0^2}^{\infty}
\frac{ d \sigma \rho_1(\sigma) (\sigma - M_0^2)}{\sigma (\sigma + Q^2)} \ .
\label{M1} 
\end{eqnarray}
In this ``M1'' model, $k_0$, $c_f$, $c_0 = M_0^2/{\Lambda}^2$,
and $c_r = M_r^2/{\Lambda}^2$ are four 
dimensionless and $C$-independent 
parameters which determine the low energy modification
of the coupling (a special case, $k_0=-1$, of M1
was presented in Ref.~\cite{Cvetic:2005my}). In general,
at high energies, this coupling differs from the
MA-coupling by $\sim {\overline \Lambda}^2/Q^2$. However, requiring
that the difference be only $\sim {\overline \Lambda}^4/Q^4$
fixes parameter $k_0$ in terms of the other three. 
Consequently, ${\overline \Lambda} \approx 0.4$ GeV
from fitting to high energy QCD observables, as in the
MA case. The remaining three parameters can be fixed by
requiring that the experimental value of $r_{\tau}$
and some other low energy observable, e.g., Bjorken
polarized sum rule $d_b(Q^2)$, be reproduced by the
afore-mentioned procedure. The experimental values
of these two observables are $r_{\tau} = 0.196 \pm 0.010$
\cite{ALEPH}
and $d_b(Q^2) =  0.16 \pm 0.11$ at $Q^2=2 \ {\rm GeV}^2$
\cite{Deur:2004ti},
where the normalization was chosen such that
$r_{\tau} = a_{\rm pt} + {\cal O}(a_{\rm pt}^2)$
and similarly for $d_b$. The quark mass effects
are subtracted (not contained) here.

The use of the MA-coupling,
in our approach (\ref{bij})-(\ref{v1}) and with
massless first three quarks and 
${\overline \Lambda} = 0.4$ GeV, gives $d_b(Q^2=2.) \approx 0.13$ 
in v1 ($0.14$ in v2) which is acceptable,
and $r_{\tau} = 0.140$ in v1 ($0.139$ in v2) 
which is not acceptable.\footnote{
For the corresponding massless Adler function $d(Q^2)$ 
we use the scheme where in Eqs.~(\ref{v1})-(\ref{v2})
$\mathcal{O}_n = \mathcal{O}_5$ \cite{CV2}.
If taking a scheme with by one order lower
precision ($\mathcal{O}_n = \mathcal{O}_4$, $\beta_3 \mapsto 0$),
the value changes to $r_{\tau}=0.146$ in v1 ($0.145$ in v2).
Observable $r_{\tau}$ is evaluated by first evaluating
the Adler function $d(Q^2)$ for complex values 
$Q^2 = m_{\tau}^2 e^{i \theta}$ and then applying the standard
contour integration in the $Q^2$-plane.
The function $F_{\mathcal{O}}^{\mathcal{A}}(t)$ for
$d(Q^2)$ was obtained in Ref.~\cite{Neubert}. 
$F_{\mathcal{O}}^{\mathcal{A}}(t)$
for $d_b(Q^2)$ can be obtained from the known large-$n_f$
expansion of $d_b(Q^2)$ \cite{Broadhurst:1993ru}, using the technique
of Ref.~\cite{Neubert}, and the full perturbation coefficients
$c_1$ and $c_2$ from \cite{LV}.}
If we require, in model M1,
the reproduction of $r_{\tau} \approx 0.196$ and
$d_b(Q^2=2.) \approx 0.13-0.14$,
in the evaluation approach v1, with massless quarks, 
and ${\overline \Lambda} = 0.4$ GeV,
we obtain for the choice $c_0 = 2$ the values 
of $c_r \approx 0.5$ and $c_f \approx 1.7$. Changing
$c_0$ while keeping it $\sim 1$ gives us by the same procedure
different values of $c_r$ and $c_f$ to
reproduce the afore-mentioned values of $r_{\tau}$ and $d_b$.
In such cases, yet another low energy observable,
the (massless) Adler function, remains quite stable under
the variation of $c_0$. In M1 we will take $c_0 =2$,
$c_r=0.5$, $c_f=1.7$ (and ${\overline \Lambda} = 0.4$ GeV).
With these parameter values: v1 approach (\ref{v1}) gives
$r_{\tau} = 0.197$ ($0.202$ when 
${\mathcal{O}}_n = {\mathcal{O}}_4$ for $d(m_{\tau}^2 e^{i \theta})$
in Eq.~(\ref{v1})) and $d_b(Q^2=2.) \approx 0.14$; 
v2 approach (\ref{v2}) gives $r_{\tau} = 0.210$ ($0.201$ when 
${\mathcal{O}}_n = {\mathcal{O}}_4$) and
$d_b(Q^2=2.) \approx 0.14$.

Yet another, simpler, modification of the MA-model is
\begin{eqnarray}
\mathcal{A}_1^{(M2)}(Q^2)& = & \mathcal{A}_1^{({\rm MA})}(Q^2) + 
{\widetilde c}_v
\frac{ {\widetilde M}_0^2 }{ Q^2 + {\widetilde M}_0^2} \ .
\label{M2}
\end{eqnarray}
In this ``M2'' model,\footnote{
In Ref.~\cite{Alekseev}, 
power correction terms $1/(Q^2)^n$
were added to $\mathcal{A}_1^{\text{(MA)}}(Q^2)$, but with
a somewhat different motivation.}
we will take the parameter values ${\widetilde c}_v = 0.2$, 
${\widetilde c}_0 \equiv {\widetilde M}_0^2/{\Lambda}^2 \approx 0.56$
and ${\overline \Lambda} = 0.4$ GeV.
With these values, v1 approach (\ref{v1}) gives
$r_{\tau} = 0.188$ ($0.194$ when 
${\mathcal{O}}_n = {\mathcal{O}}_4$) and
$d_b(Q^2=2.) \approx 0.18$;
v2 approach (\ref{v2}) gives $r_{\tau} = 0.188$ ($0.193$ when 
${\mathcal{O}}_n = {\mathcal{O}}_4$) and
$d_b(Q^2=2.) \approx 0.19$.

\begin{figure}[htb] 
\centering
\epsfig{file=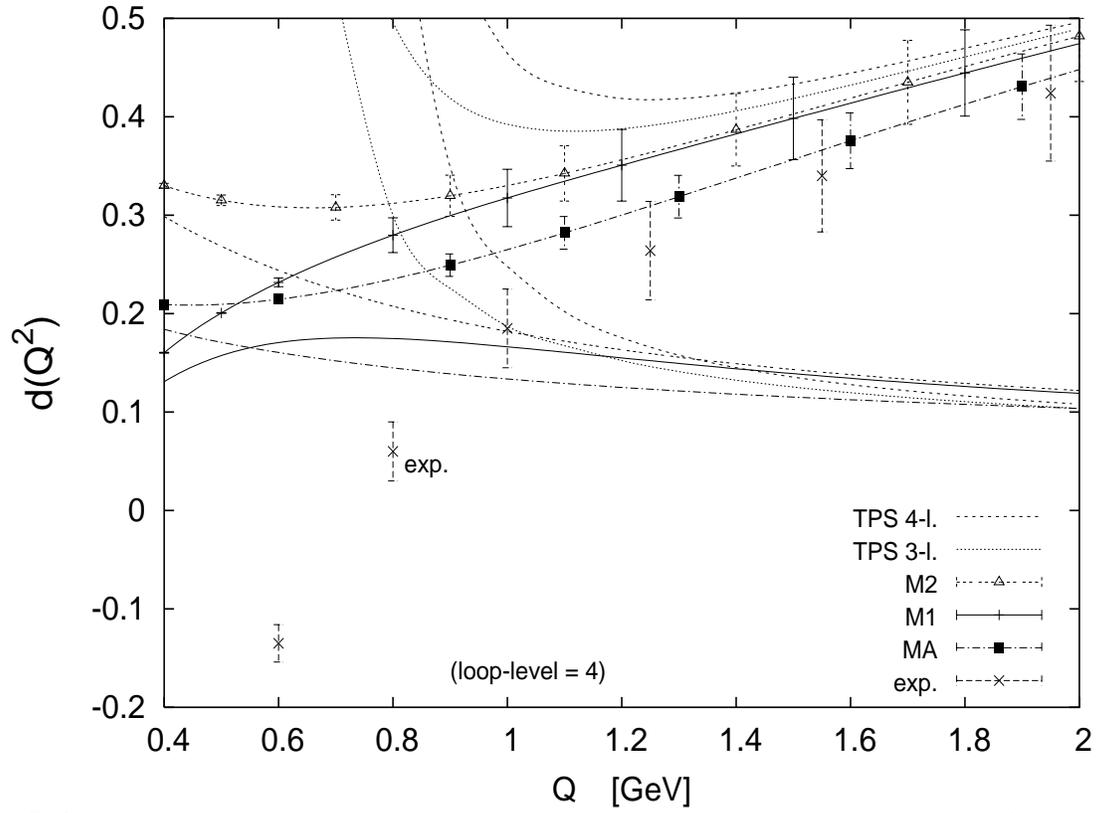,width=15.cm,height=11.cm}
\vspace{-0.4cm}
 \caption{\footnotesize  Adler function as predicted by pQCD,
and by our approach in several analytic QCD models (see the text).}
\label{dvshort}
 \end{figure}

Having fixed the parameters in the afore-mentioned models
M1 and M2, we present in Fig.~\ref{dvshort} 
low energy results of the Adler function
$d(Q^2)$ associated with the hadronic part of the electromagnetic
current, in the models MA, M1 and M2
(for a different evaluation of $d(Q^2)$, cf.~\cite{Nesterenko:2005wh}). 
The normalization was
taken again such that in the massless quark limit for $n_f=3$:
$d(Q^2) = a_{\rm pt} + {\cal O}(a_{\rm pt}^2)$.
The lower curves in Fig.~\ref{dvshort} 
represent the results of the v1-evaluation
(\ref{v1}) with three massless quarks,
in the scheme where $\mathcal{O}_n = \mathcal{O}_5$.
The higher curves represent the full quantity,
i.e., the effects of the massive $c$ and $b$ quarks
are added, with the coefficients as given in Ref.~\cite{Eidelman}
($d(Q^2) = (1/2) D(Q^2)\!-\!1$, where $D$ is defined in \cite{Eidelman}).
In the contributions of $c$ and $b$, we simply replaced $a_{\rm pt}(Q^2)$
and $a_{\rm pt}^2(Q^2)$ by $\mathcal{A}_1(Q^2)$ and
$\mathcal{A}_2(Q^2)$, and used $\Lambda = {\overline \Lambda}$
($C=-5/3$).
$\mathcal{A}_2(Q^2)$ was constructed by the fourth-order
relations (\ref{A2A3A4}).
The indicated $\pm$ uncertainties in these curves 
are those charm contributions 
which are $\propto \mathcal{A}_2(Q^2)$. 
In contrast to the massless contributions, we do not 
have yet a systematic way to analytize the massive quark
contributions.
The experimental results \cite{Eidelman} and the
truncated pQCD series were included for comparison.
Figure 1 shows that analytic versions of QCD (MA, M1, M2)
in conjunction with the skeleton-motivated approach (\ref{v1})
give results that at low energies $Q \sim 1$ GeV behave much 
better than pQCD. 

\begin{figure}[htb] 
\centering
\epsfig{file=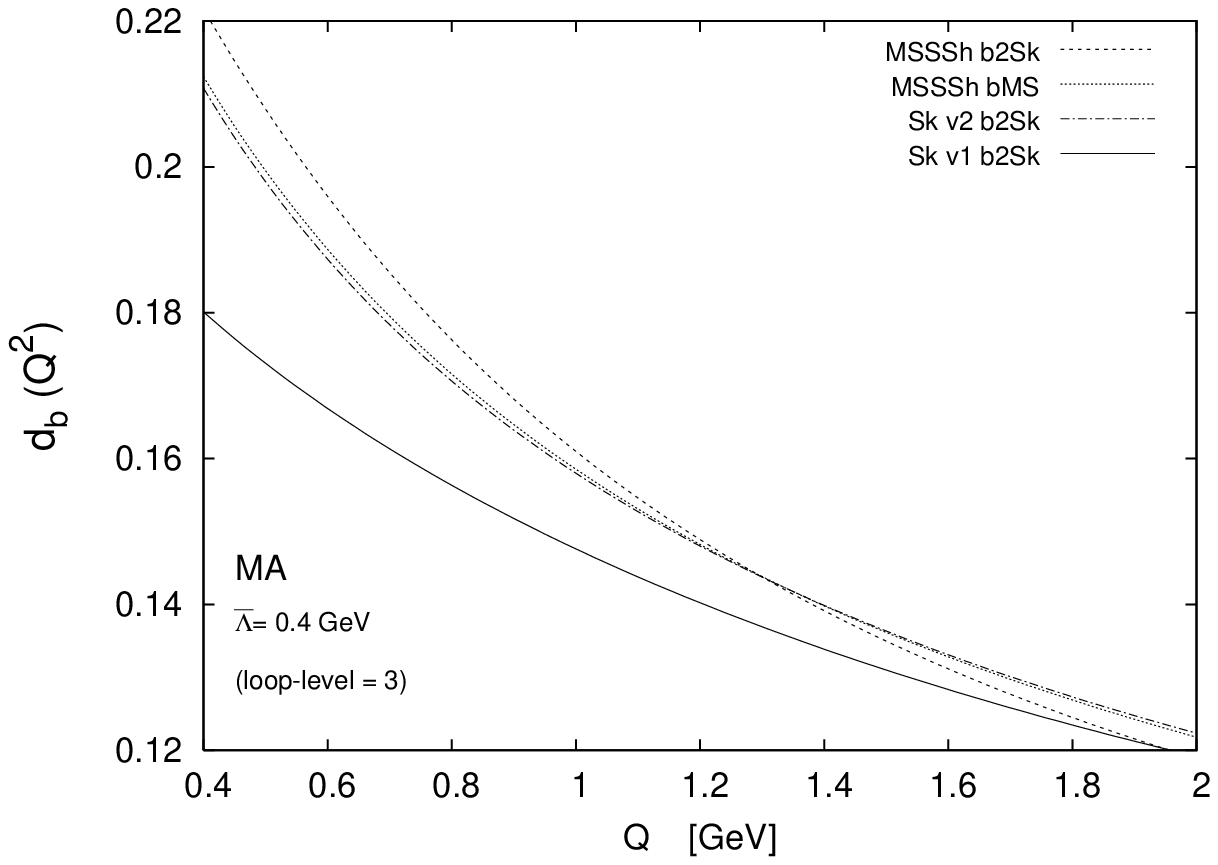,width=15.cm,height=11.cm}
\vspace{-0.4cm}
 \caption{\footnotesize Bjorken polarized sum rule
in MA as predicted by two variants of the
approach \cite{Milton:1997mi,Sh} (MSSSh),
and by our approach.}
\label{dbshort}
 \end{figure}

Application of variant 2 of our approach, Eq.~(\ref{v2}), 
gives for the Adler function results which are very close to those
of variant 1, Eq.~(\ref{v1}), because the coefficient
$s_1^{\mathcal {O}}$ is small: $s_1^{\mathcal {O}}= 1/12$. 
For the Bjorken polarized sum rule $d_b(Q^2)$, 
this is not so, because $s_1^{\mathcal {O}} = -11/12$ is appreciable.
Therefore, we will take the Bjorken sum rule as a case to look at 
numerical differences between evaluations in our approaches v1 and v2, 
Eqs.~(\ref{v1})-(\ref{v2}). In addition, we will compare
with the approach of Milton {\em et al.\/} and Shirkov 
\cite{Milton:1997mi,Sh}
(MSSSh)
\begin{equation}
\mathcal{O}_{\text{MSSSh}}(Q^2) =
\mathcal{A}_1(Q^2) + c_1 \mathcal{A}_2(Q^2) + c_2 \mathcal{A}_3(Q^2) \ .
\label{MSSSh}
\end{equation} 
In principle, the comparison between the approaches can be carried
out in any anaylitic QCD model for the coupling parameter.
However, the MSSSh approach has been applied in the
literature in the MA model.
Therefore, we carry out the comparison in this model.
In Fig.~\ref{dbshort} we present the MA-model predictions for
$d_b(Q^2)$, with various evaluations,
at third order [$\mathcal{O}_n=\mathcal{O}_4$ in 
Eqs.~(\ref{v1})-(\ref{v2})].
The results of the MSSSh approach are presented 
in two schemes: ${\overline {\rm MS}}$ (``bMS'');
and in the ``b2Sk'' scheme where $\beta_2$ 
is determined by Eqs.~(\ref{bij})
for $d_b$. The MSSSh approach uses
the three-loop MA-expressions 
$\mathcal{A}_2(Q^2)$ and $\mathcal{A}_3(Q^2)$ of
Eq.~(\ref{MAak}) (Refs.~\cite{Milton:1997mi,Sh}),
and variant 2 of our approach, Eq.~(\ref{v2}), 
uses $\mathcal{A}_2(Q^2)$ from Eqs.~(\ref{A2A3}).
Figure \ref{dbshort} shows that the evaluation of $d_b(Q^2)$
with variant 1 of our approach Eq.~(\ref{v1}) 
(``Sk v1 b2Sk'') gives, at low energies, results which differ 
significantly from the MSSSh approach. On the other hand,
variant 2 of our approach (``Sk v2 b2Sk''), i.e.,
Eq.~(\ref{v2}), gives results which are, apparently accidentally,
very close to those of the MSSSh approach
in the ${\overline {\rm MS}}$ scheme. 

In summary, we presented two variants of 
skeleton-expansion-motivated 
evaluation of observables in analytic versions of QCD,
Eqs.~(\ref{v1})-(\ref{v2}). The first variant follows more closely
the skeleton expansion, in the sense that the analytization
is performed at the beginning, in the skeleton
expansion ($a_{\rm pt} \mapsto \mathcal{A}_1$). 
In the second variant, the analytization
is performed at the end, in the form $a_{\rm pt}^k \mapsto \mathcal{A}_k$.
The second variant can be regarded
as a generalization of the evaluation approach
of Milton {\em et al.\/} and Shirkov \cite{Milton:1997mi,Sh}
(MSSSh), now including the leading-$\beta_0$ terms to all
orders in the coupling parameter. Both variants use the
formal structure of the skeleton expansion in order to divide
the original perturbation expansion into a sum of subseries
(skeleton terms), each of them renormalization scale
invariant, and then using a scheme where only the
first two subseries survive.
Further, we introduced two alternative models (M1, M2) of
analytic QCD for the coupling parameter which, for
certain values of the model parameters, reproduce
the measured values of the semihadronic $\tau$ decay
ratio.

\vspace{1.cm}

{\noindent \bf Acknowledgments}

\noindent
The authors acknowledge helpful communication with
F.~Jegerlehner.
This work was supported by Fondecyt grant 1050512 (G.C.),
Mecesup grant USA0108 and Conicyt Fellowship 3060106 (C.V.).

\begin{thebibliography}{9}

\bibitem{Neubert}
  M.~Neubert,
  Phys.\ Rev.\ D {\bf 51}, 5924 (1995);
hep-ph/9502264.

\bibitem{Bjorken:1979dk}
  J.~D.~Bjorken and S.~D.~Drell,
{\it Relativistic Quantum Fields\/}, McGraw-Hill, New York, 1965.

\bibitem{Brodsky}
  S.~J.~Brodsky, E.~Gardi, G.~Grunberg and J.~Rathsman,
  Phys.\ Rev.\ D {\bf 63}, 094017 (2001).

\bibitem{Gardi:1999dq}
  E.~Gardi and G.~Grunberg,
  JHEP {\bf 9911}, 016 (1999).

\bibitem{ShS}
  D.~V.~Shirkov and I.~L.~Solovtsov,
  hep-ph/9604363;
  Phys.~Rev.~Lett.~{\bf 79}, 1209 (1997).

\bibitem{Alekseev:2002zn}
  A.~I.~Alekseev,
  Few Body Syst.\  {\bf 32}, 193 (2003);
  J.\ Phys.\ G {\bf 27}, L117 (2001).


\bibitem{Kurashev:2003pt}
  D.~S.~Kurashev and B.~A.~Magradze,
  Theor.\ Math.\ Phys.\  {\bf 135}, 531 (2003)
  [Teor.\ Mat.\ Fiz.\  {\bf 135}, 95 (2003)];
hep-ph/0104142.

\bibitem{Nesterenko}
  A.~V.~Nesterenko,
  Phys.\ Rev.\ D {\bf 62}, 094028 (2000);
  Phys.\ Rev.\ D {\bf 64}, 116009 (2001);
  Int.\ J.\ Mod.\ Phys.\ A {\bf 18}, 5475 (2003);
  A.~V.~Nesterenko and J.~Papavassiliou,
  Phys.\ Rev.\ D {\bf 71}, 016009 (2005);
  A.~C.~Aguilar, A.~V.~Nesterenko and J.~Papavassiliou,
  J.\ Phys.\ G {\bf 31}, 997 (2005).

\bibitem{Raczka}
P.~A.~R\c{a}czka, hep-ph/0512339, presented at QCD05, Montpellier, France, July 2005.

\bibitem{Milton:1997mi}
  K.~A.~Milton, I.~L.~Solovtsov and O.~P.~Solovtsova,
  Phys.\ Lett.\ B {\bf 415}, 104 (1997);
  Phys.\ Rev.\ D {\bf 64}, 016005 (2001);
  hep-ph/0512209.

\bibitem{Bakulev:2005gw}
  A.~P.~Bakulev, S.~V.~Mikhailov and N.~G.~Stefanis,
  Phys.\ Rev.\ D {\bf 72}, 074014 (2005)
  [Erratum-ibid.\ D {\bf 72}, 119908 (2005)];
  A.~P.~Bakulev, A.~I.~Karanikas and N.~G.~Stefanis,
  Phys.\ Rev.\ D {\bf 72}, 074015 (2005).

\bibitem{CV2}
G.~Cveti\v c and C.~Valenzuela, in preparation.

\bibitem{Beneke:1992ch}
  M.~Beneke,
  Nucl.\ Phys.\ B {\bf 405}, 424 (1993).

\bibitem{Brodsky:1982gc}
  S.~J.~Brodsky, G.~P.~Lepage and P.~B.~Mackenzie,
  Phys.\ Rev.\ D {\bf 28}, 228 (1983);

\bibitem{Baikov:2002uw}
  P.~A.~Baikov, K.~G.~Chetyrkin and J.~H.~K\"uhn,
  Phys.\ Rev.\ D {\bf 67}, 074026 (2003).

\bibitem{Sh}
  D.~V.~Shirkov,
  Theor.\ Math.\ Phys.\  {\bf 127}, 409 (2001);
  Eur.\ Phys.\ J.\ C {\bf 22}, 331 (2001).

\bibitem{Cvetic:2005my}
  G.~Cveti\v c, C.~Valenzuela and I.~Schmidt,
  hep-ph/0508101, presented at QCD05, Montpellier, France, July 2005.

\bibitem{ALEPH}
  R.~Barate {\it et al.}  [ALEPH Collaboration],
  Eur.\ Phys.\ J.\ C {\bf 4}, 409 (1998);
  K.~Ackerstaff {\it et al.}  [OPAL Collaboration],
  Eur.\ Phys.\ J.\ C {\bf 7}, 571 (1999).

\bibitem{Deur:2004ti}
  A.~Deur {\it et al.},
  Phys.\ Rev.\ Lett.\ {\bf 93}, 212001 (2004);
  F.~Campanario and A.~Pineda,
  Phys.\ Rev.\ D {\bf 72}, 056008 (2005).

\bibitem{Broadhurst:1993ru}
  D.~J.~Broadhurst and A.~L.~Kataev,
  Phys.\ Lett.\ B {\bf 315}, 179 (1993).

\bibitem{LV}
S.~G.~Gorishny and S.~A.~Larin, Phys. Lett. B {\bf 172}, 109 (1986);
E.~B.~Zijlstra and W.~Van Neerven, Phys. Lett. B {\bf 297}, 377 (1992);
S.~A.~Larin and J.~A.~M. Vermaseren, Phys. Lett. B {\bf 259}, 
345 (1991).

\bibitem{Alekseev}
  A.~I.~Alekseev,
  hep-ph/0503242.

\bibitem{Nesterenko:2005wh}
  A.~V.~Nesterenko and J.~Papavassiliou,
hep-ph/0511215.

\bibitem{Eidelman}
  S.~Eidelman, F.~Jegerlehner, A.~L.~Kataev and O.~Veretin,
  Phys.\ Lett.\ B {\bf 454}, 369 (1999).

\end{thebibliography}
\end{document}